\documentclass[conference]{IEEEtran}
\usepackage{times}

\usepackage[numbers]{natbib}
\usepackage{multicol}
\usepackage[bookmarks=true]{hyperref}

\usepackage{graphicx}
\usepackage{hyperref}
\usepackage{amsmath, amssymb, amsthm}
\usepackage{siunitx}
\usepackage{placeins}
\usepackage[colorinlistoftodos]{todonotes}

\usepackage{algorithm}
\usepackage{algpseudocode}

\usepackage{accents}
\newcommand{\ubar}[1]{\underaccent{\bar}{#1}}
\newcommand\shortdots{\makebox[1em][c]{.\hss.\hss.}\thinspace}

\newcommand{\HRule}[1][\medskipamount]{\par
  \vspace*{\dimexpr-\parskip-\baselineskip+#1}
  \noindent\rule{\columnwidth}{0.5pt}\par
  \vspace*{\dimexpr-\parskip-0.5\baselineskip+#1}}
\newcommand{\boxedequation}[2]{
\begin{samepage}
\HRule[8pt]
\vspace{-4pt}
\newsec{\textbf{{#1:}}} \nopagebreak
\vspace{-.2 cm} 
#2
\HRule[8pt]
\end{samepage}
}

\newcommand{\captionSpacing}{\vspace{-5.25mm}}

\newcommand{\ie}{i.e.\ }

\usepackage{hyperref}
\usepackage{graphicx}		
\usepackage{wrapfig}
\usepackage[format=plain,font=footnotesize,labelfont=bf,labelsep=period]{caption}
\usepackage{sidecap} 
\usepackage[export]{adjustbox}

\usepackage{amsmath}
\usepackage{amssymb}
\usepackage{mathtools}
\usepackage{algorithm}
\usepackage{algpseudocode}

\usepackage{pdfsync}
\usepackage[normalem]{ulem}
\usepackage{paralist}	
\usepackage[space]{grffile} 

\usepackage{color}

\usepackage{amsthm}

\theoremstyle{definition}
\newtheorem{definition}{Definition}
\theoremstyle{remark}

\theoremstyle{definition}

\theoremstyle{definition}

\theoremstyle{definition}

\newcommand{\R}{\mathbb{R}}

\newcommand{\gap}{\vspace{.1cm}}

\newcommand{\newsec}[1]{\gap {\bf \noindent #1 }}

\definecolor{darkblue}{RGB}{0,0,102}
\definecolor{lightblue}{RGB}{77,77,148}

\definecolor{gold}{RGB}{234, 170, 0}
\definecolor{metallic_gold}{RGB}{139, 111, 78}

\renewcommand{\cal}[1]{\mathcal{ #1 }}

\newcommand{\bs}[1]{\boldsymbol{ #1 }}

\DeclareMathOperator*{\argmin}{argmin}

\begin{document}

\title{Nonlinear Model Predictive Control of Robotic Systems with Control Lyapunov Functions}




%
\author{\authorblockN{Ruben Grandia\authorrefmark{1},
Andrew J. Taylor\authorrefmark{2},
Andrew Singletary\authorrefmark{2}, 
Marco Hutter\authorrefmark{1}, and
Aaron D. Ames\authorrefmark{2}}
\authorblockA{\authorrefmark{1}ETH Z\"urich, 8092 Z\"urich, Switzerland}
\authorblockA{\authorrefmark{2}California Institute of Technology, Pasadena, CA 91125, USA}
}

\maketitle

\begin{abstract}
The theoretical unification of Nonlinear Model Predictive Control (NMPC) with Control Lyapunov Functions (CLFs) provides a framework for achieving optimal control performance while ensuring stability guarantees. In this paper we present the first real-time realization of a unified NMPC and CLF controller on a robotic system with limited computational resources. These limitations motivate a set of approaches for efficiently incorporating CLF stability constraints into a general NMPC formulation. We evaluate the performance of the proposed methods compared to baseline CLF and NMPC controllers with a robotic Segway platform both in simulation and on hardware. The addition of a prediction horizon provides a performance advantage over CLF based controllers, which operate optimally point-wise in time. Moreover, the explicitly imposed stability constraints remove the need for difficult cost function and parameter tuning required by NMPC. Therefore the unified controller improves the performance of each isolated controller and simplifies the overall design process. 
\end{abstract}

\IEEEpeerreviewmaketitle

\section{Introduction}

\FloatBarrier 
\begin{figure}[t]
    \centering
\begin{minipage}[t]{0.49\columnwidth}
    \centering
   \includegraphics[width=0.75\textwidth]{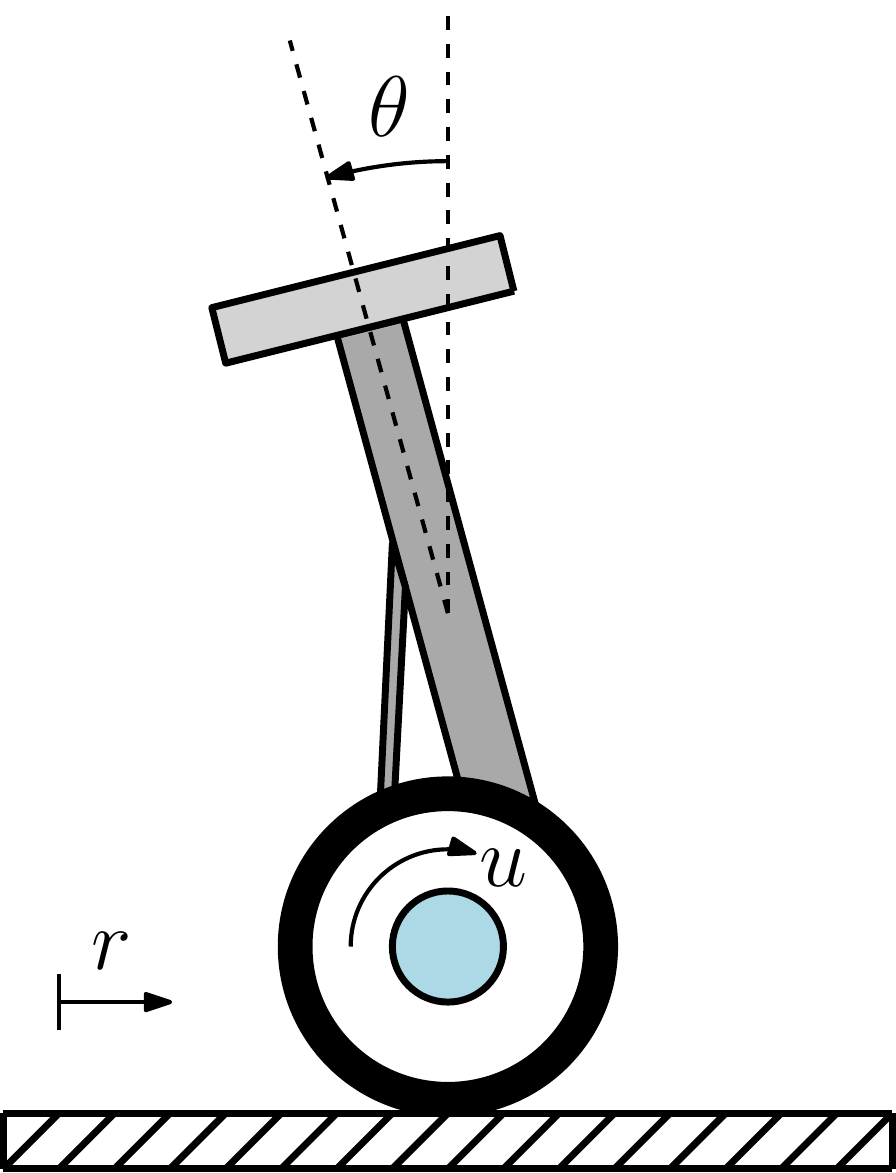}
\end{minipage}%
\begin{minipage}[t]{0.49\columnwidth}
    \centering 
   \includegraphics[width=0.90\textwidth]{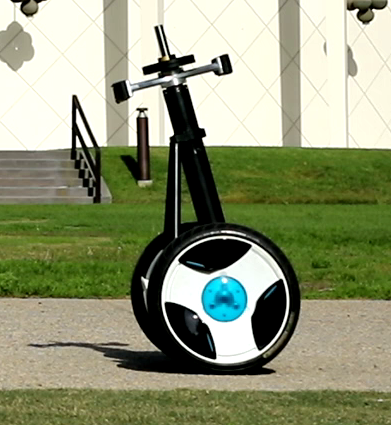}
\end{minipage}%
    \caption{Left: Segway model for simulation and control design. Right: Physical Segway system in outdoor experiment environment.}
    \label{fig:segway-outside}
    \captionSpacing
\end{figure}

Deploying autonomous and versatile robots into the real world comes with the challenge of ever increasing complexity in sensing, decision making, and actuation. One difficulty in designing controllers for such complex systems lies in the need to simultaneously meet a large set of design requirements. Achieving stable and safe behavior is often in conflict with performance objectives, and finding the right balance between these requirements can be a challenging task. 

A disjoint, hierarchical approach is typically taken in this context. High-level trajectories are planned to satisfy performance objectives via computationally intensive nonlinear optimization, and local feedback controllers are separately designed to ensure stability. The goal of this paper is to directly integrate these two components by constructing controllers that simultaneously optimize performance along a horizon and satisfy local stability constraints in a computationally efficient manner. In particular, we seek to unify guarantee-based methods of nonlinear control with the optimization-based view in model predictive control. To this end, we combine the guarantees endowed by Control Lyapunov Functions (CLFs) \cite{artstein1983stabilization, sontag1989universal} with the optimal performance of Nonlinear Model Predictive Control (NMPC) \cite{Bock1984, Mayne2000, Rawlings2017}. 

Lyapunov methods are a powerful tool for certifying stability properties of nonlinear systems \cite{Khalil}. The use of Control Lyapunov Functions to synthesize stabilizing controllers for robotic platforms has become increasingly popular \cite{galloway2015torque, ma2017bipedal, nguyen2015optimal}, often via quadratic programs (QPs) \cite{ames2014rapidly, ames2013towards}. Despite the optimization-based formulation of these controllers, they often fail to achieve long-term optimal behavior. This deficiency arises due to the fact that the cost of these optimization problems fails to incorporate the future behavior of the system, but is instead point-wise optimal \cite{freeman1996inverse}.

In contrast, Nonlinear Model Predictive Control (NMPC) emphasizes performance by solving a finite horizon optimal control problem online and applying the first element of the computed open-loop input trajectory to the system \cite{Rawlings2017}. This optimization is repeatedly solved with each newly measured state to obtain the next control input trajectory. While this class of control laws can achieve strong performance in practice \cite{garcia1989model, qin2003survey, Grune2017}, and allows intuitive specification of the desired behaviour, additional assumptions must be met to certify closed-loop stability. In classical discrete-time NMPC, stability is guaranteed by an appropriately designed terminal penalty and terminal constraint \cite{Mayne2000, CHEN1998, Grune2017}.

The integration of Lyapunov methods with NMPC is not a new idea. Lyapunov methods have been used to construct stabilizing terminal conditions \cite{Jadbabaie2001}, or to analyze stability in the absence thereof \cite{Jadbabaie2005}. Another approach incorporates the stability condition required by a CLF along the prediction horizon found with NMPC \cite{Primbs2000, Yu2001}. As noted in \cite{Primbs2000}, this approach has several desirable properties such as the absence of a terminal cost, stability for any horizon length, and recovery of the CLF-QP \cite{ames2014rapidly} or infinite horizon optimal controllers when considering the limiting behavior of the finite horizon. The idea of imposing stability constraints along the horizon has appeared in other forms such as contractive state constraints \cite{Oliveira2000}, and has been applied within the context of chemical process control \cite{Mahmood2014, Wu2018}, economic cost functions \cite{Heidarinejad2012}, and switched nonlinear systems \cite{Mhaskar2005}.

While this existing work has analyzed the stability and optimality properties obtained through the unification of CLFs and NMPC, there has been little attention to the practical and computational aspects of the resulting nonlinear optimization problem. Limited computational resources and fast system dynamics present a challenge to the deployment of these unified methods to modern robotic systems. Indeed, to the best of our knowledge, such a control scheme has not yet been applied to robotic systems experimentally. 

To achieve the goal of experimental realizability, we develop a new methodology for combining CLFs with NMPC. We describe practical methods to efficiently solve the resulting nonlinear optimization problems and ultimately realize the proposed controllers in simulation and, for the first time, experimentally on a robotic platform; in this case, a Segway hardware platform seen in Fig.~\ref{fig:segway-outside}. While each proposed method provides theoretical stability guarantees, significant differences in computational efficiency and performance are observed. Furthermore, we find that the pairing of these control methodologies leads to improved performance over CLF methods and significantly reduced tuning of prediction horizon length and terminal conditions for NMPC methods.

Our paper is organized as follows. Section \ref{sec:background} provides a review of CLFs and the stability guarantees they yield, and reviews the NMPC problem and how it is solved in practice. In Section \ref{sec:method} we propose a set of methods for incorporating CLF stability constraints into the NMPC problem, and provide additional details on implementation. Lastly, in Section \ref{sec:results} we provide results from both simulation and hardware that demonstrate the ability of this unified control approach to achieve stability and improve performance.

\section{Background}
\label{sec:background}
In this section we provide background information on Control Lyapunov Functions (CLFs) and Nonlinear Model Predictive Control (NMPC). This information supports the specific framework unifying CLFs and NMPC in Section \ref{sec:method}. 

\subsection{Control Lyapunov Functions}
\label{sec:clfs}
Consider a state space $\cal{X}\subset\R^n$ and a control input space $\cal{U}\subset\R^m$, where it is assumed $\cal{X}$ is path-connected and $0\in\cal{X}$. Consider the control-affine dynamic system given by:
\begin{equation}
    \dot{x} = f(x) + g(x)u.
    \label{eq:AffineDynamics}
\end{equation}
where $x\in\cal{X}$, $u\in\cal{U}$, and $f:\cal{X}\to\R^n$ and $g:\cal{X}\to\R^{n\times m}$ are Lipschitz continuous on $\cal{X}$. Further assume that $f(0)=0$, or that the origin is an equilibrium point of the system. As in \cite{Khalil}, we define a class $\cal{K}$ function as a continuous function $\alpha:[0,a)\to\R_+$, with $a>0$, $\alpha(0)=0$ and $\alpha$ strictly monotonically increasing (denoted $\alpha\in\cal{K}$).  If $a=\infty$ and $\lim_{r\to\infty}\alpha(r)=\infty$, then $\alpha$ is said to be a \textit{class $\cal{K}_\infty$ function} ($\alpha\in\cal{K}_\infty$). This type of function can be interpreted as a type of nonlinear gain function, noting the linear gain function $\alpha(r) = kr$ with $k>0$ satisfies this definition. Given this definition, we define Control Lyapunov Functions (CLFs) as in \cite{artstein1983stabilization, lin1991universal}.

\begin{definition}[\it{Control Lyapunov Functions}] 
\label{def:CLF}
A continuously differentiable function $V:\cal{X}\to\R_+$ is a \textit{Control Lyapunov Function (CLF)} for \eqref{eq:AffineDynamics} on $\cal{X}$ if there exists $\alpha_1,\alpha_2,\alpha_3\in\cal{K}_\infty$ such that for all $x\in\cal{X}$:
\begin{align}
\label{eq:posdef}
    \alpha_1(\Vert x\Vert) \leq V(x) & \leq \alpha_2(\Vert x\Vert) \\ \label{eq:clf} \inf_{u\in\cal{U}}\dot{V}(x,u) & \leq -\alpha_3(\Vert x\Vert),
\end{align}
\end{definition}
This definition can be restated with $\alpha_1, \alpha_2, \alpha_3\in\cal{K}$ with resulting stability guarantees holding locally. The existence of a CLF for \eqref{eq:AffineDynamics} implies the existence of a continuous (except possibly at $x=0$) state-feedback controller $k:\cal{X}\to\cal{U}$ that renders the origin globally asymptotically stable \cite{artstein1983stabilization, sontag1989smooth}. It is possible to make $k$ continuous at $x=0$ if $V$ satisfies the small-control property \cite{sontag1989universal}. If the functions $\alpha_1,\alpha_2,\alpha_3$ take the form $\alpha_i(r) = c_ir^2$, $i=1,2,3$, the resulting stability is global exponential stability, with the magnitude of the state upper bounded by a function exponentially decaying in time:
\begin{equation}
    \Vert x(t) \Vert \leq M\Vert x(0) \Vert e^{-\gamma t}
\end{equation}
with $M,\gamma>0$. Similarly, the CLF can be upper bounded:
\begin{equation}
    V(x(t)) \leq V(x(0))e^{-\gamma t}
    \label{eq:V_upperbound}
\end{equation}
This preceding bound will be useful for enforcing Lyapunov stability guarantees within the discrete time NMPC problem.

The CLF definition implies the existence of a point-wise set of stabilizing control inputs:
\begin{equation}
\label{eq:clfUset}
    \cal{U}_{\textrm{clf}}(x) = \{ u\in\cal{U} ~\vert~ \dot{V}(x,u)\leq-\alpha_3(\Vert x\Vert) \}.
\end{equation}
Thus a CLF characterizes a stabilizing feedback controller as a controller $k:\cal{X}\to\cal{U}$ such that $k(x)\in\cal{U}_{\textrm{clf}}(x)$ for all $x\in\cal{X}$. Furthermore, upon selection of such a controller, the CLF is a certificate of stability for the closed loop system:
\begin{equation}
    \dot{x} = f(x) + g(x)k(x).
\end{equation}
Establishing that a given function $V$ serves as a CLF for \eqref{eq:AffineDynamics} is often done for robotic systems constructively by specifying a controller taking values in $\cal{U}_{\textrm{clf}}(x)$ for all $x\in\cal{X}$ \cite{taylor2019episodic, kolathaya2019pd}. Note that for any $x\in\cal{X}$ the set $\cal{U}_{\textrm{clf}}(x)$ is described by an affine inequality in $u$ due to the affine nature of the dynamics:
\begin{equation}
    \dot{V}(x,u) = \frac{\partial V}{\partial x}(x) \left( f(x) + g(x) u \right).
\end{equation}
Due to this, the CLF itself may then be used to synthesize a optimization-based controller with more desirable properties using quadratic programs \cite{ames2014rapidly, ames2013towards, galloway2015torque}. Specifically, we obtain a feedback control law $k(x)$ that satisfies the inequality \eqref{eq:clf}:
\begin{align}
\label{eq:CLF-QP}
k(x) =  \,\,\underset{u \in \cal{U}}{\argmin}  &  \quad \frac{1}{2} u^\top R u + q^\top u + p \\
\mathrm{s.t.} \quad & \quad \frac{\partial V}{\partial x}(x)\left(f(x) +g(x)u \right)
\leq -\alpha_3(\Vert x\Vert)) \nonumber
\end{align}
where $R$ is positive definite and $\cal{U}$ assumed to be a polytope. Feasibility of this optimization problem is guaranteed by the satisfaction of the constraint \eqref{eq:clf} and Lipschitz continuity of this controller has been studied in \cite{ morris2015continuity,jankovic2018robust}.

This controller is point-wise optimal \cite{freeman1996inverse}, and takes a greedy approach to specifying control inputs. This often leads to poor performance compared to even non-optimization based controllers as there is no consideration of the future behavior of the system when the input is chosen. In addition to challenges in achieving longer horizon optimality, these controllers face difficulty in implementation on robotic platforms. The stability guarantees endowed by these controllers assume a continuous-time implementation, which is not possible on many modern digital control systems. Instead, control inputs are chosen and held for a small interval of time in a zero-order-hold manner. Lyapunov stability of zero-order-hold systems has been studied utilizing an approximate discretization of the nonlinear dynamics \cite{Nesic1999, Nesic2004}, or in the context of model predictive control \cite{Gyurkovics2004,nevsic2006receding,Mhaskar2006, Grune2007}.

\subsection{Nonlinear Model Predictive Control}
\label{sec:nmpc}
Nonlinear Model Predictive Control (NMPC) offers an alternative to CLF-based methods for controlling nonlinear systems, and is inherently designed to resolve the challenges of longer horizon optimality at the expensive of  higher online computational cost.

We consider a \textit{direct NMPC} approach to transform the continuous optimal control problem into a finite dimensional nonlinear program (NLP) \cite{Bock1984}. The continuous control signal $u(t)$ is parameterized over subintervals of the prediction horizon $[0,T]$ to obtain a finite dimensional decision problem. This creates a fixed grid of nodes $k \in \{0, \shortdots, N\}$ defining control times $t_k$ separated by intervals of duration $\delta t = T/(N-1)$. In this work, we consider a piecewise constant, or zero-order-hold, parameterization of the input. Denoting $x_k = x(t_k)$ and integrating the continuous dynamics in \eqref{eq:AffineDynamics} over an interval leads to a discrete time representation of the dynamics:
\begin{equation}
   x_{k+1} = f^d(x_k, u_k) = x_k + \int_{t_k}^{t_k+\delta t} f(x(\tau)) + g(x(\tau))u_k \text{ d}\tau.
   \label{eq:discrete-dynamics}
\end{equation}
The integral in \eqref{eq:discrete-dynamics} is numerically approximated with an integration method of choice to achieve a desired approximation accuracy of the evolution of the continuous time system under the zero-order-hold commands.

The general NMPC problem presented below can be formulated by defining and evaluating a cost function and constraints on the grid of nodes. Here, we write the problem in parametric form, depending on the current measured state $\hat{x}$ and additional parameters contained in $p$, and with a subset of the constraints implemented as soft-constraints with slack terms:
\begin{subequations}
\label{eq:NMPC} %
\begin{flalign}
	& \mathrlap{\underset{\begin{subarray}{c}
		X, U, S
		\end{subarray}}{\min}\,\, l_N(x_N, p) + \phi(s_N)  + \sum_{k=0}^{N-1} l_k(x_k, u_k, p) + \phi(s_k)} \\
	\label{eq:NMPC-ic}
	&\quad\text{s.t.}&  x_0 - \hat{x} &= 0, \\
	\label{eq:NMPC-dyn}
	&& x_{k+1} - f^d_k(x_k, u_k) &= 0, &&k = 0,\shortdots,N\!-\!1,\\
	&& h_k(x_k, u_k, p) &\leq s_k, &&k = 0,\shortdots,N\!-\!1, \\
	\label{eq:NMPC-hk}
	&& h_N(x_N, p) &\leq s_N, \\
	\label{eq:NMPC-hN}
	&& s_k &\geq 0, &&k = 0,\shortdots,N,
\end{flalign} 
\end{subequations}
where $X = [x_0^\top, \dots x_N^\top]^\top$, $U = [u_0^\top, \dots u_{N-1}^\top]^\top$, and $S = [s_0^\top, \dots s_N^\top]^\top$ are the sequences of state, input, and slack variables respectively. The nonlinear cost and constraint functions $l_k$, $h_k$, $l_N$, and $h_N$, are allowed to vary depending on the node index $k$ and are dependent on problem specific parameters $p$ and the current measured state $\hat{x}$. The slack variables are penalized with $\phi(s) = z^\top s + \frac{1}{2} s^\top Z s $, an exact $\ell_1-\ell_2$ penalty \cite{scokaert1999feasibility}. Collecting all decision variables into a vector, $w = [X^\top, U^\top, S^\top]^\top$, problem \eqref{eq:NMPC} can be framed as a general nonlinear program (NLP):
\begin{equation}
    \underset{w}{\min} \quad F(w, p) \quad  \text{s.t.} \quad
     \left\{
    \begin{array}{rl}
        G(w, p) & = 0, \\
        H(w, p) & \leq 0.
    \end{array}
\right.
\label{eq:NLP}
\end{equation}

\subsection{Sequential Quadratic Programming (SQP)}
Interior-Point Methods (IP) and Sequential Quadratic Programming (SQP) are two popular families of algorithms for solving general NLPs \cite{NoceWrig06}. Additionally, the sparsity of \eqref{eq:NLP} induced by the underlying structure of \eqref{eq:NMPC} can be exploited to obtain solutions in real-time and at a high sampling rate, which is necessary in many dynamic robotic applications. An overview of recent advances in sparsity exploiting algorithms and software tools is provided in \cite{Kouzoupis2018}.

SQP approaches offer a distinct advantage in that successive problem instances may be warm-started with solutions from preceding instances. This serves to further decrease computation time as it is often only feasible to take a single SQP step per control iteration \cite{Li1989}. As NMPC computes optimal control inputs over a horizon, successive instances of \eqref{eq:NMPC} are similar and portions of the preceding optimal control sequence can be use to warm-start the following iteration, enabling convergence across multiple iterations of the problem, rather than iterating until convergence on one instance of the problem \cite{Diehl2002}.

SQP based methods apply Newton-type iterations to KKT optimality conditions for \eqref{eq:NLP}, assuming some regularity conditions on the constraints \cite{Mangasarian1967}. The Lagrangian of the NLP in \eqref{eq:NLP} is defined as:
\begin{equation}
    \mathcal{L}(w, \lambda, \mu, p) = F(w, p) + \lambda^\top G(w, p)  + \mu^\top H(w, p),
    \label{eq:NLP_lagrangian}
\end{equation}
with Lagrange multipliers $\lambda$, and $\mu \geq 0$ corresponding to equality and inequality constraints, respectively. The Newton iterations can be equivalently computed by solving the following potentially non-convex QP \cite{NoceWrig06}:
\begin{subequations}
\label{eq:SQP_QPSubproblem}
\begin{align}
	\underset{\begin{subarray}{c}
		\delta w
		\end{subarray}}{\min} & \quad
	 \nabla_w F(w_i, p)^\top \delta w + \frac{1}{2} \delta w^\top B_i \delta w  \label{eq:SQP-qp-cost} & \\
	\quad\text{s.t} & \quad  G(w_i, p) + \nabla_w G(w_i, p)^\top \delta w = 0, & \label{eq:SQP-qp-eqconstr}  \\
	& \quad H(w_i, p) + \nabla_w H(w_i, p)^\top \delta w \leq 0, & \label{eq:SQP-qp-ineqconstr}
\end{align}
\end{subequations}
where the decision variables, $\delta w = w - w_i$, define the update step relative to the current iteration $w_i$, and the Hessian $B_i = \nabla^2_w\mathcal{L}(w_i,\lambda_i,\mu_i,p)$. Computing the solution to \eqref{eq:SQP_QPSubproblem} provides a decision variable update, $\delta w_i$, and updated Lagrange multipliers $\lambda^{QP}_i$ and $\mu^{QP}_i$. 
These iterations are ran until the variables $w_i$, $\lambda_i$, and $\mu_i$ converge. This iterative approach is summarized in Algorithm \ref{alg:SQP}.

\begin{algorithm}[tb]
\caption{Sequential Quadratic Programming (SQP)}
\label{alg:SQP}
\begin{algorithmic}
\State \textbf{Given} $p, w_0, \lambda_0, \mu_0, F, G, H$ 
\State \textbf{Initialize} $(i, w_i, \lambda_i, \mu_i) \gets (0, w_0, \lambda_0, \mu_0)$
\While {NotConverged($w_i, \lambda_i, \mu_i$) }
    \State compute $\nabla_w F(w_i, p), B_i, H(w_i, p), \nabla_w H(w_i, p),$ \\ \qquad\qquad\qquad $ G(w_i, p), \nabla_w G(w_i, p).$\\
    \State $(\delta w_i, \lambda_{i}^{QP}, \mu_{i}^{QP}) \gets $ Solve \eqref{eq:SQP_QPSubproblem}
    \State $w_{i+1} \gets  w_i + \delta w_i $
    \State $\lambda_{i+1} \gets \lambda_{i}^{QP}$
    \State $\mu_{i+1} \gets \mu_{i}^{QP} $
    \State $i \gets i + 1 $
\EndWhile
\end{algorithmic}
\end{algorithm}
Convergence of the SQP algorithm leads to a state and input sequence, $X^\star$ and $U^\star$, respectively. The first element of $U^\star$, denoted as $u_0$, can be applied to the system, after which the SQP algorithm is run again to determine a new control input sequence. The application of SQP as a subroutine within a NMPC feedback controller is provided in Algorithm \ref{alg:SQP-NMPC}.

\section{Unifying CLFs with NMPC}
\label{sec:method}
In this section we explore different ways of integrating the stability based CLF-QP and performance driven NMPC controllers discussed in Section \ref{sec:background}. These different methods will be evaluated experimentally in Section \ref{sec:results}.

The NMPC framework presented in Algorithm \ref{alg:SQP-NMPC} can be interpreted as a closed-loop feedback controller, $k_{\textrm{NMPC}}:\cal{X}\to\cal{U}$. As described in Section \ref{sec:clfs}, for each state $x\in\cal{X}$, a CLF defines a point-wise set of stabilizing control inputs $\cal{U}_{\textrm{clf}}(x)$ given in \eqref{eq:clfUset}. To inherit the stability guarantees provided by the CLF, we need to restrict the NMPC controller to these stabilizing inputs, such that $k_{\textrm{NMPC}}(x)\in\cal{U}_{\textrm{clf}}(x)$ for all $x\in\cal{X}$. 

As noted in \cite{Primbs2000}, if only the first input in the input sequence is applied before the input sequence is recomputed, the restriction of the NMPC controller to stabilizing inputs can be achieved by directly imposing the CLF condition only on the first input, subject to the current measured state $\hat{x}$:
\begin{equation}
    h_{CLF}(\hat{x}, u_0) = \frac{\partial V}{\partial x}(\hat{x})\left(f(\hat{x}) +g(\hat{x})u_0 \right) \leq -\alpha_3(\Vert \hat{x}\Vert).
    \label{eq:h_CLF}
\end{equation}
\begin{algorithm}[tb]
\caption{SQP - NMPC}
\label{alg:SQP-NMPC}
\begin{algorithmic}
\State \textbf{Given} $w^0, \lambda^0, \mu^0$
\State \textbf{Initialize} $(j, w^j, \lambda^j, \mu^j) \gets (0, w^0, \lambda^0, \mu^0)$
\While {ControllerIsRunning()}
    \State $\hat{x} \gets$ StateEstimation()
    \State $y_{ref} \gets$ Commands() 
    \State $p \gets (\hat{x}, y_{ref}$)
    \State $(w^{j+1}, \lambda^{j+1}, \mu^{j+1}) \gets $ Solve SQP($p, w^j, \lambda^j, \mu^j$)
    \State $U^\star \gets$ ExtractInputSequence($w^{j+1}$)
    \State $u_0 \gets$ ExtractFirstInput($U^\star$)
    \State $f(x)+g(x)u \gets$ ApplyInput($u_0$)
    \State $j \gets j + 1 $
\EndWhile
\end{algorithmic}
\end{algorithm}As in the case of the controller \eqref{eq:CLF-QP}, for a given state, this constraint is affine in the decision variable $u_0$. Due to this, the SQP subroutine \eqref{eq:SQP_QPSubproblem} contains the CLF constraint without approximation, and will therefore, in the same way as the CLF-QP, compute a stabilizing control input after solving just one QP. In the context of dynamic robotic platforms, this provides an advantage over general NMPC in that it does not require the computational cost of multiple Newton iterations to converge to a potentially stabilizing control input.

Beyond the constraint on the first input, any further modifications to the NMPC problem are done explicitly to increase performance or accommodate the discrete time implementation of control inputs. In the following, we will discuss additional constraints that serve to increase performance beyond that of the controller given by \eqref{eq:CLF-QP} while achieving stability.

\subsection{Extended Horizon Constraints}
While the preceding CLF constraint \eqref{eq:h_CLF} enforces the selection of a stabilizing control input, the resulting theoretical stability properties rely on the controller being applied continuously in time, as noted in Section \ref{sec:clfs}. As this is not possible in practice, it desirable to incorporate additional stability constraints that enforce stable behavior when control is implemented in a zero-order-hold fashion. NMPC is an advantageous framework for these types of constraints as future states to reflect desired stability properties. 

In particular, consider the bound in time on the Lyapunov function established with exponential stability in \eqref{eq:V_upperbound}. While this bound is continuous in time, comparison principles \cite{Khalil} may be used to formulate an analogous discrete time bound at the $k$th node:
\begin{equation}
\label{eq:h_LSS}
    h_{LLS}(x_k, \hat{x}) = V(x_k) - V(\hat{x})e^{-\gamma k \cdot \delta t} \leq 0.
\end{equation}

The constraints given by $h_{CLF}$ and $h_{LLS}$ can be combined in varying ways along the length of the prediction horizon. The basic approach, denoted \textit{CLF-0} and presented in \eqref{eq:clf-NMPC-cost} to \eqref{eq:clf-NMPC-CLF-0}, implements the $h_{CLF}$ constraint only at the initial node.
This is extended in the \textit{CLF-All} approach, where the $h_{CLF}$ constraint is enforced at each node in the horizon in \eqref{eq:clf-NMPC-CLF-All}. These constraints are no longer affine since at nodes $k \geq 1$ both state $x_k$ and input $u_k$ are decision variables, and are non-linearly coupled in \eqref{eq:h_CLF}.

In the approach denoted \textit{LLS-N}, we enforce the $h_{CLF}$ constraint at the first node and enforce the $h_{LLS}$ constraint at only the final node in the prediction horizon in \eqref{eq:clf-NMPC-LLS-N}.
A similar bound, which relies on evaluating $V$ at the final node after the system has been simulated under a control law given by Sontag's universal formula \cite{sontag1989smooth}, is enforced in \cite{Primbs2000}. Our bound differs in that it is controller independent and only relies on the bound from exponential stability. Lastly, in the approach denoted \textit{LLS-All}, the level set constraint $h_{LLS}$ is applied at each node in \eqref{eq:clf-NMPC-LLS-N}, forming an exponentially contracting funnel along the horizon. Additionally, for all formulations, the inputs are bounded $\ubar{u} \leq u_k \leq \bar{u}$, enforcing $u_k\in\cal{U}$.

\boxedequation{CLF-NMPC}{
\begin{subequations}
\label{eq:clf-NMPC}
\begin{flalign}
	& \mathrlap{\underset{\begin{subarray}{c}
		X, U, S
		\end{subarray}}{\min}\quad \phi(s_N)  + \sum_{k=0}^{N-1} \frac{1}{2} u_k^\top u_k + \phi(s_k)} \label{eq:clf-NMPC-cost}\\
	\label{eq:clf-NMPC-ic}
	&\quad\text{s.t}&  x_0 - \hat{x} &= 0, \\
	\label{eq:clf-NMPC-dyn}
	&& x_{k+1} - f^d_k(x_k, u_k) &= 0, &&k = 0,\shortdots,N\!-\!1,\\
	&& \ubar{u} \leq u_k &\leq \bar{u}, &&k = 0,\shortdots,N\!-\!1, \\
	&& s_k &\geq 0, &&k = 0,\shortdots,N, \\
	&\mathrlap{\textit{CLF-0}:}& h_{CLF}(\hat{x}, u_0) &\leq s_0, 
	\label{eq:clf-NMPC-CLF-0}\\ 
	&\mathrlap{\textit{Additionally, for}} \nonumber \\
	&\mathrlap{\textit{CLF-ALL}:}& h_{CLF}(x_k, u_k)  &\leq s_k, \ &&k = 1,\shortdots,N\!-\!1, 
	\label{eq:clf-NMPC-CLF-All}\\
	&\mathrlap{\textit{LLS-N}:}& h_{LLS}(x_N, \hat{x})  &\leq s_N, 
	\label{eq:clf-NMPC-LLS-N}\\
	&\mathrlap{\textit{LLS-All}:}& h_{LLS}(x_k, \hat{x}) & \leq s_k, &&k = 1,\shortdots,N.
	\label{eq:clf-NMPC-LLS-All}
\end{flalign}
\end{subequations}
}

\subsection{Quadratic Approximation Strategy}
When applying the SQP algorithm presented in Section~\ref{sec:nmpc} to the nonlinear formulations presented in \eqref{eq:clf-NMPC}, the quadratic subproblem \eqref{eq:SQP_QPSubproblem} is repeatedly solved. As we seek to deploy NMPC on dynamic robotic platforms, it is critical that these optimization problems are well conditioned and do not provide difficulty to numerical solvers. In particular, when $B_i$ in \eqref{eq:SQP-qp-cost} is positive semi-definite (p.s.d), the resulting QP is convex and can be efficiently solved \cite{Kouzoupis2018,stellato2018osqp}.

To ensure this, an approximate (p.s.d) Hessian can be used instead of the full Hessian of the Lagrangian. For \eqref{eq:clf-NMPC}, the objective function has a least-squares form, \ie $F(w, p) = \frac{1}{2} \| R(w, p) \|^2 $, in which case the Gauss-Newton approximation,
\begin{equation}
  B_i \approx \nabla_w R(w_i, p)^\top \nabla_w R(w_i, p),  
  \label{eq:GaussNewton}
\end{equation}
proves effective in practice \cite{bock1983recent,houska2011auto}. This neglects the curvature of $R(w, p)$, as well as the contribution by the curvature of the constraints. We use this strategy for the \textit{CLF-0} and \textit{CLF-All} formulations. 

For the \textit{LLS-N} and \textit{LLS-All} formulations, we retain the contribution of the LLS constraints in the Hessian approximation.
The second term in \eqref{eq:h_LSS} is independent of the decision variables, and the properties of this constraint are thus directly determined by the structure of the CLF $V$. In particular, if the CLF used is convex (as is the case for many constructive techniques for producing CLFs \cite{taylor2019episodic, kolathaya2019pd}), then
the approximation
\begin{align}
 B_i \approx \nabla_wR(w_i,p)^\top&\nabla_wR(w_i,p) \nonumber\\ &+ \sum_{k} \mu_{k,LLS}\nabla_w^2 h_{k,LLS},\label{eq:GaussNewtonLLS}
\end{align}
remains positive definite, and, as we will show in Section~\ref{sec:NumericalResults}, improves convergence compared to a Gauss-Newton approximation.

\subsection{Baseline Comparisons}
To understand how unifying these two control methodologies impacts performance and stability, it is necessary to compare against baseline controllers given by both \eqref{eq:CLF-QP} and \eqref{eq:NMPC}. In particular, elements should be shared between controllers to limit the impact of tuning on performance. To this end, we begin by synthesizing a CLF using feedback-linearization based constructive techniques discussed in \cite{taylor2019episodic}, which enables the consideration of underactuated systems. 

Consider an output $y:\cal{X}\to\R^k$ with relative degree 2 \cite{Sastry99} and $k\leq m$, a time-varying reference trajectory $y:\R_+\to\R^k$, and define the tracking error $e:\cal{X}\times\R_+\to\R^k$: \begin{equation}
    e(x,t) = y(x) - y_d(t).
\end{equation} A feedback-linearizing controller exists, $k_{\textrm{fbl}}:\cal{X}\to\cal{U}$ that yields linear closed-loop error dynamics given by:
\begin{equation}
    \dot{\eta}(x,t) = A\eta(x,t), \quad \eta = \begin{bmatrix}e(x,t) \\ \dot{e}(x,t) \end{bmatrix} 
\end{equation}
where the eigenvalues of $A\in\R^{2k\times2k}$ have negative real part. For any positive definite $Q\in\R^{2k\times2k}$, the Continuous Time Lyapunov Equation (CTLE):
\begin{equation}
    A^\top P + PA = -Q
\end{equation}
has a positive definite solution $P\in\R^{2k\times2k}$. This enables the synthesis of a quadratic (and convex) Lyapunov function $V:\cal{X}\times\R_+\to\R$ given by $V(x,t) = \eta(x,t)^\top P\eta(x,t)$ with time derivative $\dot{V}(x,t) = -\eta(x,t)^\top Q\eta(x,t)$, which is negative definite. Furthermore, the existence of the feedback linearizing controller $k_{\textrm{fbl}}$ implies that $V$ is a CLF, as:
\begin{equation}
    \inf_{u\in\cal{U}}\dot{V}(x,t,u)  \leq -\lambda_{\textrm{min}}(Q)\Vert \eta(x,t)\Vert_2^2.
\end{equation}
This CLF can be used in the following \textit{CLF-QP} controller that achieves exponential stability with $\gamma = \frac{\lambda_{\textrm{min}}(Q)}{\lambda_{\textrm{max}}(P)}$. The constraint is slacked for numerical conditioning with $z,Z\geq0$.
\boxedequation{CLF-QP}{
\begin{align}
\label{eq:clfbase}
\underset{u \in \cal{U},s\in\R_+}{\min}  &  \quad \frac{1}{2} u^\top u + zs + \frac{1}{2}Zs^2  \\
\mathrm{s.t.} \quad & \quad \dot{V}(x,t,u)
\leq -\lambda_{\textrm{min}(Q)}\Vert \eta(x,t)\Vert_2^2+s \nonumber
\end{align}
}

To synthesize a baseline \textit{NMPC-$\beta$} controller, elements from the construction of the CLF can be utilized. In particular, $Q$ can be used as a running cost on the state, and the terminal value of the CLF can penalized as in \cite{Jadbabaie2001}:
\boxedequation{NMPC-$\bs{\beta}$}{
\begin{subequations}
\label{eq:NMPC-beta}
\begin{flalign}
	& \mathrlap{\underset{\begin{subarray}{c}
		X, U
		\end{subarray}}{\min}\,\,\, \beta V(x_N,t_N) \!+\! \sum_{k=0}^{N-1} \eta(x_k,t_k)^\top Q \eta(x_k,t_k) \!+\! \frac{1}{2} u_k^\top u_k } \label{eq:NMPC-beta-cost}\\
	\label{eq:NMPC-beta-ic}
	&\quad\text{s.t}&  x_0 - \hat{x} &= 0, \\
	\label{eq:NMPC-beta-dyn}
	&& x_{k+1} - f^d_k(x_k, u_k) &= 0, &&k = 0,\shortdots,N\!-\!1,\\
	&& \ubar{u} \leq u_k &\leq \bar{u}, &&k = 0,\shortdots,N\!-\!1, 
\end{flalign}
\end{subequations}
}
That is, the CLF is used as a terminal cost and scaled up with the parameter $\beta$. As noted in \cite{Rawlings2017}
if $\beta$ is selected large enough, stability can be achieved without the need to specify a terminal state constraint. The baseline NMPC problem has constraints on the initial condition and dynamic evolution given by \eqref{eq:NMPC-ic} and \eqref{eq:NMPC-dyn}, but does not include any CLF-based constraints. The inputs are also constrained such that $u\in\cal{U}$.

\section{Simulation \& Experimental Results}
\label{sec:results}

In this section we provide details on the implementation of the methods established in Section \ref{sec:method} on a Segway platform, and discuss simulation and experimental results.

\subsection{Segway System \& Implementation}

Dynamics simulations provide an environment for assessing attainable levels of performance of the various approaches. The simulated dynamics model reflects a modified Ninebot E+ Segway platform, seen in Fig.~\ref{fig:segway-outside}. We consider a planar representation of the Segway, with state $x = \begin{bmatrix}r & \theta & \dot{r} & \dot{\theta} \end{bmatrix}^\top\in\R^4$ where $r$ is the horizontal position and $\theta$ is the pitch angle. The input to the Segway, $u=\begin{bmatrix} i \end{bmatrix}\in[-20,20]$, is current to the Segway motors. The equations of motion are derived via the Newton-Euler equations for an asymmetric, two-wheeled inverted pendulum with torque input. The asymmetry of the system leads to an unforced equilibrium at $x_e = \begin{bmatrix} 0, \theta_e, 0, 0 \end{bmatrix}$ with $\theta_e = 0.138$. In the NMPC controllers, a forward Euler time discretization is used with $\delta t = \SI{0.01}{\second}$.

State estimation on the physical Segway is done with wheel encoders and IMU data from a VectorNav VN-100. All computations are performed on board on an ARM Cortex-A57 (quad-core) @ 2GHz CPU running the ERIKA3 RTOS. For each NMPC formulation, all functions, gradients, and Hessians in \eqref{eq:SQP_QPSubproblem} are found using the CasADi auto-differentiation framework \cite{Andersson2019}. This leads to a QP with a fixed sparsity pattern, which we solve with the sparsity-exploiting solver OSQP \cite{stellato2018osqp}. We solve a single QP per control iteration, unless otherwise stated.

\subsection{Simulation Results}
To compare the behavior of the different control approaches, we considered a stabilization task. In particular, we simulated the system under each controller with a fixed initial condition and an objective of stabilizing to the unforced equilibrium point $x_e$. The performance of each controller was quantified by the average input norm over a 2 second time horizon, which provides an assessment of the total input used. These averages appear in Table~\ref{tab:min-norm-sim}.

\begin{table}[h]
\centering
\caption{Average input norm along the simulation horizon against prediction horizon length (N) for different controller formulations defined in Section~\ref{sec:method}. Absences indicate failure to stabilize the system.}
\label{tab:min-norm-sim}
\begin{tabular}{c c c c c c c}
\hline
\textit{N} & 1 & 10 & 20 & 30 & 40 & 50  \\ \hline

\textit{CLF-QP}                       & 1.085 & 1.085   & 1.085          & 1.085 & 1.085 & 1.085 \\

\textit{CLF-0}                      & 1.085 & 1.085   & 1.085          & 1.085          & 1.085 & 1.085 \\ 
\textit{CLF-All}                      & 1.085 & 1.072   & 0.952          & 0.849          & \textbf{0.794} & \textbf{0.769} \\ 
\textit{LLS-N}                      & 1.083 & 0.957   & 0.889          & 0.842          & 0.808 & 0.784 \\ 
\textit{LLS-All}                      & 1.083 & 0.956   & 0.887 & 0.839          & 0.805 & 0.782          \\

\textit{NMPC-0.1}                      & - & -   & 3.232 & 2.435          & 2.036 & 1.783    \\ 

\textit{NMPC-1} & - & 3.026 & 2.019 & 1.732 & 1.574 & 1.471 \\

\textit{NMPC-10} & \textbf{0.828} & \textbf{0.607} & \textbf{0.704} & \textbf{0.823} & 0.926 & 1.006 \\ \hline
\end{tabular}
\end{table}

We see that the \textit{CLF-QP} and \textit{CLF-0} controllers select identical inputs over the entire trajectory and have the largest average input of any CLF-based controllers. This indicates that the addition of a prediction horizon to the baseline \textit{CLF-QP} controller only improves its performance if the stability $h_{CLF}$ constraint is applied further along the horizon. We also see that with no horizon ($N\!=\!1$), all of the CLF-based controllers recover similar performance to the baseline \textit{CLF-QP}. If we consider the controllers that impose constraints further along the horizon we see that their average input consistently decreases with horizon length, with the \textit{CLF-All} controller marginally outperforming the \textit{LLS} controllers at the longest horizon lengths. 

The performance of the baseline \textit{NMPC}-$\beta$ controllers heavily depended on the weighting of the terminal cost. At shorter horizons the \textit{NMPC-0.1} and \textit{NMPC-1} controllers failed to stabilize the system, and saw improved performance as the horizon grew longer. In contrast, the \textit{NMPC-10} controller demonstrated the best performance of all controllers at shorter horizons, but saw worsening performance as the horizon grew longer. This illustrates the issues that arise when both stability and performance are achieved through the cost, rather than decoupled through constraints as in the CLF-based formulations.

\begin{figure}[tb]
    \centering
   \includegraphics[width=\columnwidth]{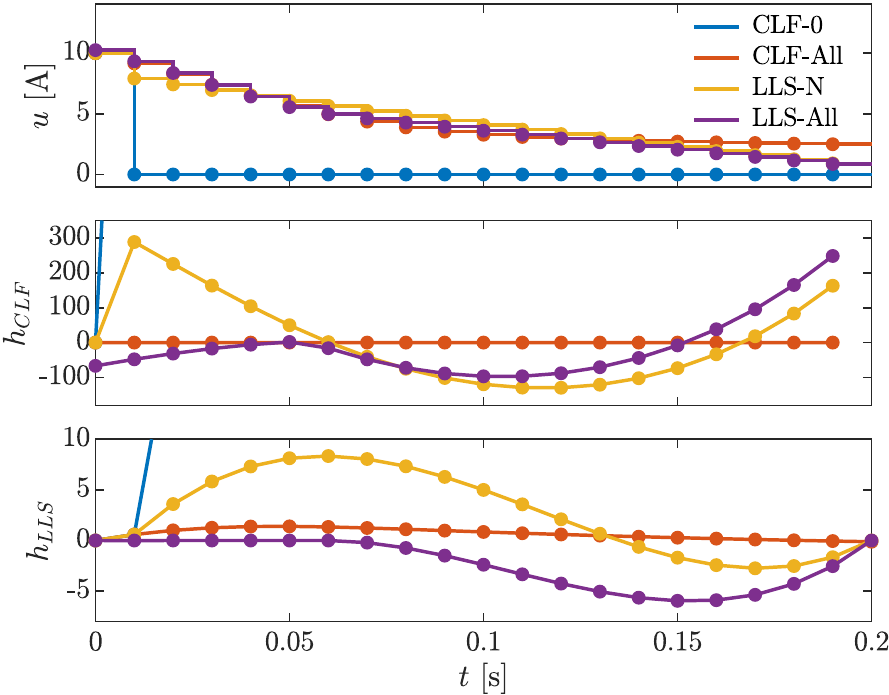}
    \caption{Initial \textit{CLF-NMPC} solutions stabilizing to $x_e$ for $N=20$ and $\hat{x} = [0, \frac{\pi}{8}, 0, 0]^\top$. 
    \textbf{Top:} The optimal input sequence determined by each controller, with the \textit{CLF-0} controller dropping to zero beyond the first node. \textbf{Middle:} Evaluation of the bound on $\dot{V}$ in \eqref{eq:h_CLF}. The controllers with the $h_{CLF}$ constraint meet this bound at all required points, while the two \textit{LLS} controllers violated it at various points along the trajectory. \textbf{Bottom:} Evaluation of the the level set bound in \eqref{eq:h_LSS}. This bound is satisfied at all necessary points by the \textit{LSS} controllers, but is violated by the \textit{CLF-All} controller due to the zero-order-hold implementation.}
    \label{fig:InitialSolutions}
    \captionSpacing
\end{figure}

To more clearly understand the possible behaviors of the various \textit{CLF-NMPC} controllers, we visualize the solutions to each controller obtained at the initial condition, \ie $\hat{x}=x(0)$ in Fig.~\ref{fig:InitialSolutions}. As the \textit{CLF-0} controller is only required to satisfy the stability constraint at the initial node, its input quickly drops to zero and the $h_{CLF}$ constraint is violated in the next step in the horizon. In contrast, the \textit{CLF-All} controller meets the $h_{CLF}$ constraint along the entirety of the horizon as required. Despite meeting this constraint on $\dot{V}$ at each node, the \textit{CLF-All} controller slightly fails to meet the implied level-set bounds along the horizon. This is due to the fact that the constraint is checked only at the beginning of each interval and is not required to hold over the interval which the control input is held over.

The \textit{LLS} controllers both satisfy the level constraints at the required points, with the \textit{LLS-N} controller violating the level set bounds earlier in the horizon. Despite meeting these level set bounds and the required points, both controllers violate the associated bound on $\dot{V}$, with the \textit{LLS-All} controller satisfying the $h_{LLS}$ constraint loosely early in the trajectory before satisfying it tightly at the end of the trajectory.

\begin{figure}[tb]
    \centering
   \includegraphics[width=\columnwidth]{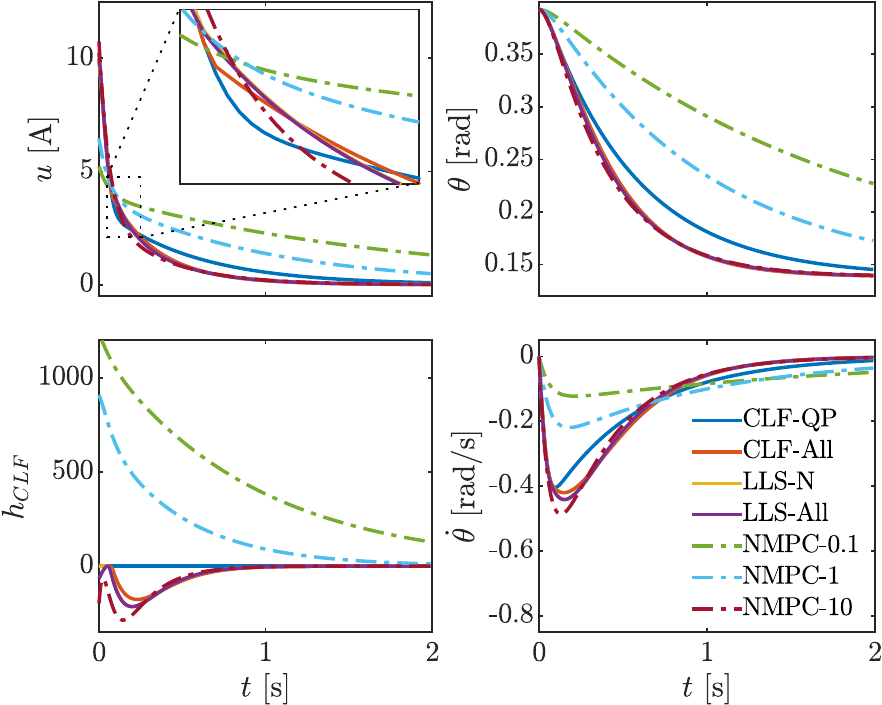}
    \caption{Simulation results stabilizing to $x_e$ for $x(0) = [0, \frac{\pi}{8}, 0, 0]^\top$. The evaluation of the CLF constraint \eqref{eq:h_CLF} along the simulation horizon is shown in the bottom-left plot. The system input, pitch angle, and pitch rate are shown in the top-left, top-right, and bottom-right respectively. The lower weighted \textit{NMPC}-$\beta$ controllers take significantly more time to converge to the equilibrium point.}
    \label{fig:SimulationPlots}
    \captionSpacing
\end{figure}

The evolution of the system under these controllers with a horizon length $N=30$ is captured in Fig.~\ref{fig:SimulationPlots}. We see that all CLF-based controllers satisfy the stability constraint \eqref{eq:h_CLF} during the entire simulation. The most significant difference in the behavior of these controllers arises in the magnitude of the input taken early in the trajectory. The best performing controllers applying higher inputs earlier in the trajectory to stabilize the system quickly and avoid accruing input over more of the trajectory.

\begin{figure}[tb]
    \centering
   \includegraphics[width=\columnwidth]{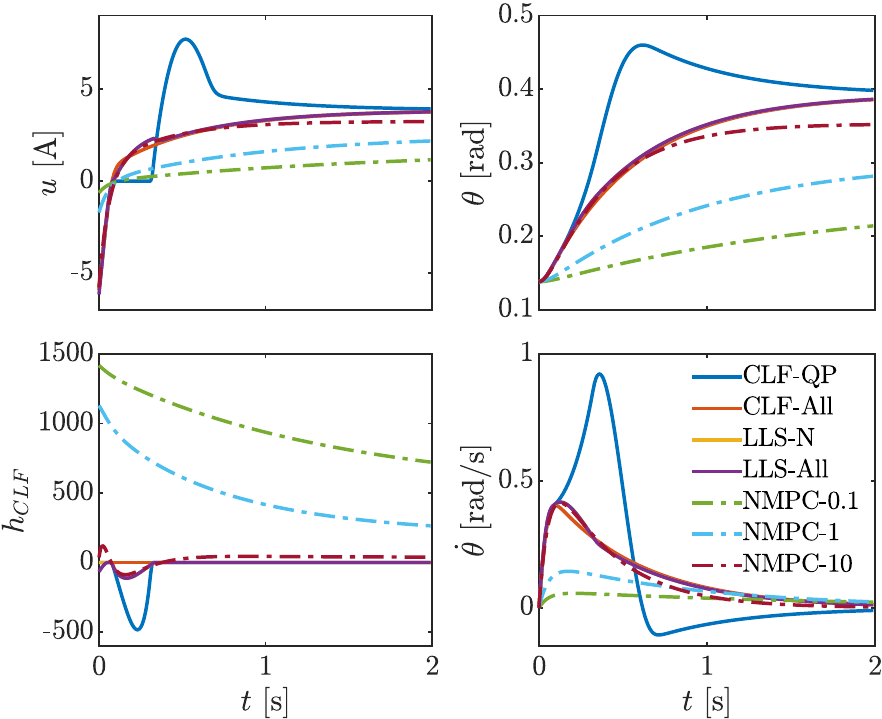}
    \caption{
    Simulation results stabilizing to $x_d = [0, \frac{\pi}{8}, 0, 0]^\top$ from $x^\star$. The evaluation of the CLF constraint \eqref{eq:h_CLF} along the simulation horizon is shown in the bottom-left plot. The system input, pitch angle, and pitch rate are shown in the top-left, top-right, and bottom-right respectively. The \textit{CLF-QP} controller overshoots the desired state, while the \textit{CLF-NMPC} controllers slowly approach it. The \textit{NMPC}-$\beta$ controllers do not converge due to the trade-off between input and state error in the cost function.}
    \label{fig:SimulationBackwardsPlots}
    \captionSpacing
\end{figure}

We additionally perform a simulation in the opposite direction, driving the Segway from its unforced equilibrium $x_e$ to a desired state, $x_d = [0.0, \frac{\pi}{8}, 0.0, 0.0]^\top$. The resulting trajectories are seen in Fig.~\ref{fig:SimulationBackwardsPlots}. In this scenario we see that the \textit{CLF-QP} controller demonstrates significantly different behavior from the \textit{CLF-NMPC} controllers. In particular, the \textit{CLF-QP} controller takes an initial input to drive the system towards the desired state, and then allows gravity to carry it to this point. This results in a large overshoot past the desired state, where as the \textit{CLF-NMPC} controllers approach the equilibrium point more slowly.

The \textit{NMPC-$\beta$} controllers fail to converge to the desired state. This arises due to the fact that the cost on the input is not centered about the input that makes the desired state an equilibrium point, \ie $\vert u-u_{ref}\vert_2^2$. Because both the input and the state error appear in the cost, larger error is accepted in the state to minimize the input. This highlights the flexibility in choosing the cost function in the \textit{CLF-NMPC} formulations. General cost functions that do not necessarily obtain their minimum at the goal can be used, which opens up the possibility of using economic cost functions \cite{Heidarinejad2012}. 
 
\subsection{Numerical Results}
\label{sec:NumericalResults}
To understand the feasibility of deploying these control approaches on the computationally limited Segway platform, we investigated the convergence rate of each formulation on a single instance of the NMPC problem. We also considered variants of the \textit{LLS} formulations using the Gauss-Newton approximation in \eqref{eq:GaussNewton} (denoted $\textit{LLS-N}^{GN}$ and $\textit{LLS-All}^{GN}$) and compare them to using the modified Hessian in \eqref{eq:GaussNewtonLLS}.

\begin{figure}[tb]
    \centering
   \includegraphics[width=0.95\columnwidth]{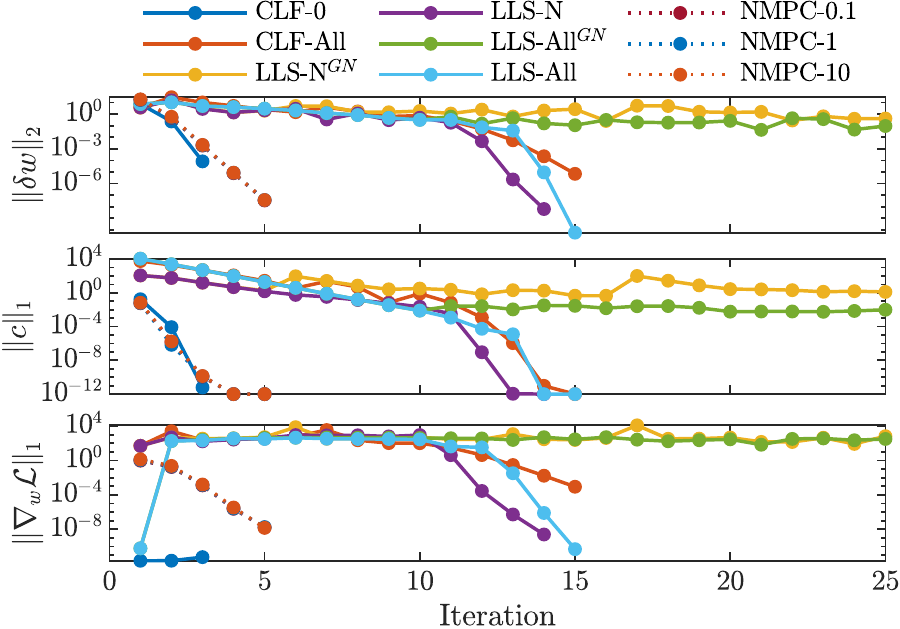}
    \caption{Convergence of Algorithm~\ref{alg:SQP} applied to a cold-started NMPC problem with $\hat{x} = x_d = [0, \frac{\pi}{8}, 0, 0]^\top$. \textbf{Top:} Step size, $\|\delta w\|_2$, \textbf{Middle:} Constraint violation, $\|c(w)\|_1$, \textbf{Bottom:} First order optimality condition, $\| \nabla_w \cal{L} \|_1$. The \textit{CLF-0} and \textit{NMPC}-$\beta$ methods converge quickly, while the \textit{CLF-NMPC} methods take longer. The \textit{LLS} formulations without the modified Hessian in \eqref{eq:GaussNewtonLLS} stop progressing beyond a certain point.}
    \label{fig:ConvergencePlots}
    \captionSpacing
    \vspace{-8mm}
\end{figure}

We execute Algorithm~\ref{alg:SQP} for each formulation until the constraints are sufficiently satisfied, $\|c(w)\|_1 \leq 10^{-6}$, and no further progress is made in cost, $|F(w_i) - F(w_{i-1})| \leq 10^{-6}$. The optimization is fully cold-started such that all decision variables and Lagrange multipliers are initialized to zero.
The problem is set up with $\hat{x} = x_d = [0, \frac{\pi}{8}, 0, 0]^\top$ and a horizon length of $N=30$. The step size, constraint satisfaction, and first order optimality are plotted in Fig.~\ref{fig:ConvergencePlots}.

The \textit{CLF-0} and \textit{NMPC-$\beta$} formulations converge rapidly as they have a quadratic cost function and affine constraints. The only nonlinearity in these problems therefore arises from the system dynamics, making the QP subproblem a good model for the full problem. The \textit{CLF-All}, \textit{LLS-N}, and \textit{LLS-All} have nonlinear constraints along the horizon, and therefore take significantly longer to converge. Nonetheless, the convergence rate accelerates over the last few iterations, indicating that these formulations can still perform well when starting close to the solution. Finally, the \textit{LLS-N$^{GN}$}, and \textit{LLS-All$^{GN}$} formulations initially decrease the constraint violation, but fail to make further progress after a certain point. 
\subsection{Experimental Results}

\begin{figure}[tb]
    \centering
   \includegraphics[width=0.95\columnwidth]{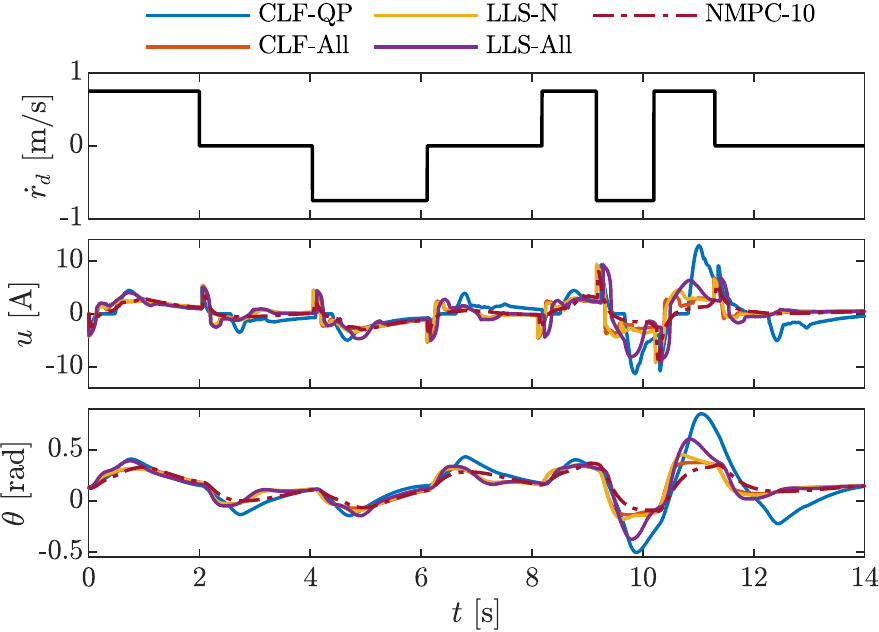}
    \caption{Experimental results from trajectory tracking. \textbf{Top:} Desired velocity profile, \textbf{Middle:} Input profile, \textbf{Bottom:} Pitch angle profile. The \textit{CLF-QP} controller displays more aggressive behavior due to no prediction horizon, while the \textit{NMPC-1} controller fails to stabilize the system.}
    \label{fig:HardwarePlots}
    \captionSpacing
    \vspace{-3mm}
\end{figure}

In addition to simulation, we also demonstrate the ability of this unified control approach on the physical Segway in Fig.~\ref{fig:segway-outside}. In particular, we define a desired angular trajectory in time:
\begin{equation}
    \theta_d(x, t) = \theta_e - K_v (\dot{r} - \dot{r}_{d}(t) ),
\end{equation}
where $\dot{r}_{d}(t)$ is a commanded velocity and $K_v = 0.3$ is a velocity feedback gain, leading to error dynamics given by $e(x,t) = \theta-\theta_d(x, t)$. These dynamics are used to synthesize a quadratic CLF as per Section \ref{sec:method}. This CLF was used to formulate a \textit{CLF-QP} controller given by \eqref{eq:clfbase}, a \textit{NMPC-1} and \textit{NMPC-10} controller with cost given by \eqref{eq:NMPC-beta-cost}, and \textit{CLF-All}, \textit{LLS-N}, and \textit{LLS-All} controllers. Each controller was used to track the desired trajectory and then stabilize the system. The results of these experiments can be seen in Fig.~\ref{fig:HardwarePlots}. We see that all controllers except the \textit{NMPC-1} controller are able to stabilize the system. The \textit{CLF-QP} displays aggressive behavior compared to the \textit{NMPC} controllers as it does not incorporate a prediction horizon. 
The average input and control frequency along the experimental horizon is seen in Table~\ref{tab:min-norm-exp}. We see that at a horizon of $N=30$ the \textit{NMPC-10} controller has the best performance, with the \textit{NMPC-1} controller omitted due to failure to stabilize. Of the \textit{CLF-NMPC} controllers the \textit{CLF-All} controller has the best performance. We note that although the baseline \textit{NMPC-10} controller outperforms the proposed methods, this required tuning of the cost function and matches the behavior seen in simulation at this horizon length. We see that the \textit{CLF-NMPC} methods have a higher computational cost than the \textit{NMPC-10} controller. This follows as the NLP has additional constraints related to stability that must be met. In that sense, \textit{LLS-N} is an appealing approach among the \textit{CLF-NMPC} methods, as it imposes only two stability constraints.

\begin{table}[h]
\caption{Average input norm ($\bar{u}$) and computation time ($t_{CPU}$) in \SI{}{\milli\second} along the experiment horizon with prediction horizon $N=30$ for the different controller formulations defined in Section~\ref{sec:method}.}
\centering
\begin{tabular}{ c c c c c c}
\hline
 & CLF-QP & CLF-All & LLS-N & LLS-All &  NMPC-10  \\ \hline
\multicolumn{1}{l|}{$\bar{u}$} & 2.081     &    1.594     &   1.666    &    1.898     &     \textbf{1.152} \\   
\multicolumn{1}{l|}{$t_{CPU}$} & 1.25 & 5.56 & 4.17 & 6.13 & 3.11 \\
 \hline
\end{tabular}
\label{tab:min-norm-exp}
\vspace{-4mm}
\end{table}

\section{Conclusion} 
\label{sec:conclusion}
In conclusion, we have presented a novel set of approaches for unifying CLFs and NMPC on robotic platforms with limited computational resources. The use of a SQP algorithm with modified Hessian was proposed to efficiently solve the resulting nonlinear optimization problem. The different unified formulations were analyzed in simulation, for the first time demonstrated on hardware, and were shown to improve performance beyond baseline CLF and NMPC methods. In particular for forced equilibria, the CLF-NMPC method converges without modifications while the cost-driven baseline NMPC does not. Furthermore, the unified methods all achieved stability where as the stability baseline NMPC methods was sensitive to cost function parameters. As system complexity increases, such manual tuning becomes increasingly difficult. In this space, we see an opportunity for the presented CLF-NMPC methods, where stability is explicitly embedded and requires no further tuning.

\section*{Acknowledgments}
R.Grandia and M. Hutter are supported via the European Union’s Horizon 2020 research and innovation programme under grant agreement No 780883. A. Taylor, A. Singletary, and A. Ames are supported via DARPA awards HR00111890035 and NNN12AA01C, and NSF awards 1923239 and 1924526.



\clearpage
\addtolength{\textheight}{-9.25cm}
\bibliographystyle{plainnat}
\bibliography{main}

\end{document}